\documentclass[letterpaper,twocolumn,english,manuscript,aip]{revtex4-1}
\usepackage[T1]{fontenc}
\usepackage{graphicx}

\makeatletter

\pdfpageheight\paperheight
\pdfpagewidth\paperwidth

\@ifundefined{textcolor}{}
{%
 \definecolor{BLACK}{gray}{0}
 \definecolor{WHITE}{gray}{1}
 \definecolor{RED}{rgb}{1,0,0}
 \definecolor{GREEN}{rgb}{0,1,0}
 \definecolor{BLUE}{rgb}{0,0,1}
 \definecolor{CYAN}{cmyk}{1,0,0,0}
 \definecolor{MAGENTA}{cmyk}{0,1,0,0}
 \definecolor{YELLOW}{cmyk}{0,0,1,0}
 }

\makeatother

\usepackage{babel}

\begin{document}

\title{Do topological models provide good information about electricity
infrastructure vulnerability?}

\author{Paul Hines}

\email{paul.hines@uvm.edu}

\author{Eduardo Cotilla-Sanchez}

\email{eduardo.cotilla-sanchez@uvm.edu}

\affiliation{School of Engineering, University of Vermont, Burlington, VT 05405,
USA}

\author{Seth Blumsack}

\email{blumsack@psu.edu}

\affiliation{Department of Energy and Mineral Engineering, Pennsylvania State
University, University Park, PA 16802, USA}
\begin{abstract}
In order to identify the extent to which results from topological
graph models are useful for modeling vulnerability in electricity
infrastructure, we measure the susceptibility of power networks to
random failures and directed attacks using three measures of vulnerability:
characteristic path lengths, connectivity loss and blackout sizes.
The first two are purely topological metrics. The blackout size calculation
results from a model of cascading failure in power networks. Testing
the response of 40 areas within the Eastern US power grid and a standard
IEEE test case to a variety of attack/failure vectors indicates that
directed attacks result in larger failures using all three vulnerability
measures, but the attack vectors that appear to cause the most damage
depend on the measure chosen. While our topological and power grid
model results show some trends that are similar, there is only a mild
correlation between the vulnerability measures for individual simulations.
We conclude that evaluating vulnerability in power networks using
purely topological metrics can be misleading.\end{abstract}
\maketitle
\begin{quotation}
Electricity infrastructures are vital to the operation of modern society,
yet they are notably vulnerable to cascading failures. Understanding
the nature of this vulnerability is fundamental to the assessment
of electric energy reliability and security. A number of articles
have recently used topological (graph theoretic) models to assess
vulnerability in electricity systems. In this article we illustrate
that under some circumstances these topological models can lead to
provocative, but ultimately misleading conclusions. We argue that
emperical comparisons between topological models and higher fidelity
models are neccessary in order to draw firm conclusions about the
utility of complex networks methods.
\end{quotation}

\section{Introduction}

Motivated by the importance of reliable electricity infrastructure,
numerous recent papers have applied complex networks methods \cite{Albert:2002,Boccaletti:2006}
to study the structure and function of power grids. Results from these
studies differ greatly. Some measure the topology of power grids and
report exponential degree distributions \cite{Amaral:2000,Albert:2004,Hines:2010a},
whereas others report power-law distributions \cite{Barabasi:1999,Chassin:2005}.
Some models of the North American power grid suggest that power grids
are more vulnerable to directed attacks than to random failures \cite{Albert:2004,Holmgren:2006},
even though power grids differ from from scale-free networks in topology.
Recently, Wang and Rong \cite{Wang:2009} used a topological model
of cascading failure and argue that attacks on nodes (buses) transporting
smaller amounts of power can result in disproportionately large failures.
Albert et al. \cite{Albert:2004} draw the opposite conclusion using
similar data. Because of the potential implications of these results
for infrastructure security, these papers \cite{Albert:2004,Wang:2009}
have attracted the attention of government and media \cite{Markoff:2010}.

\begin{figure}[tbh]
\includegraphics[width=1\columnwidth]{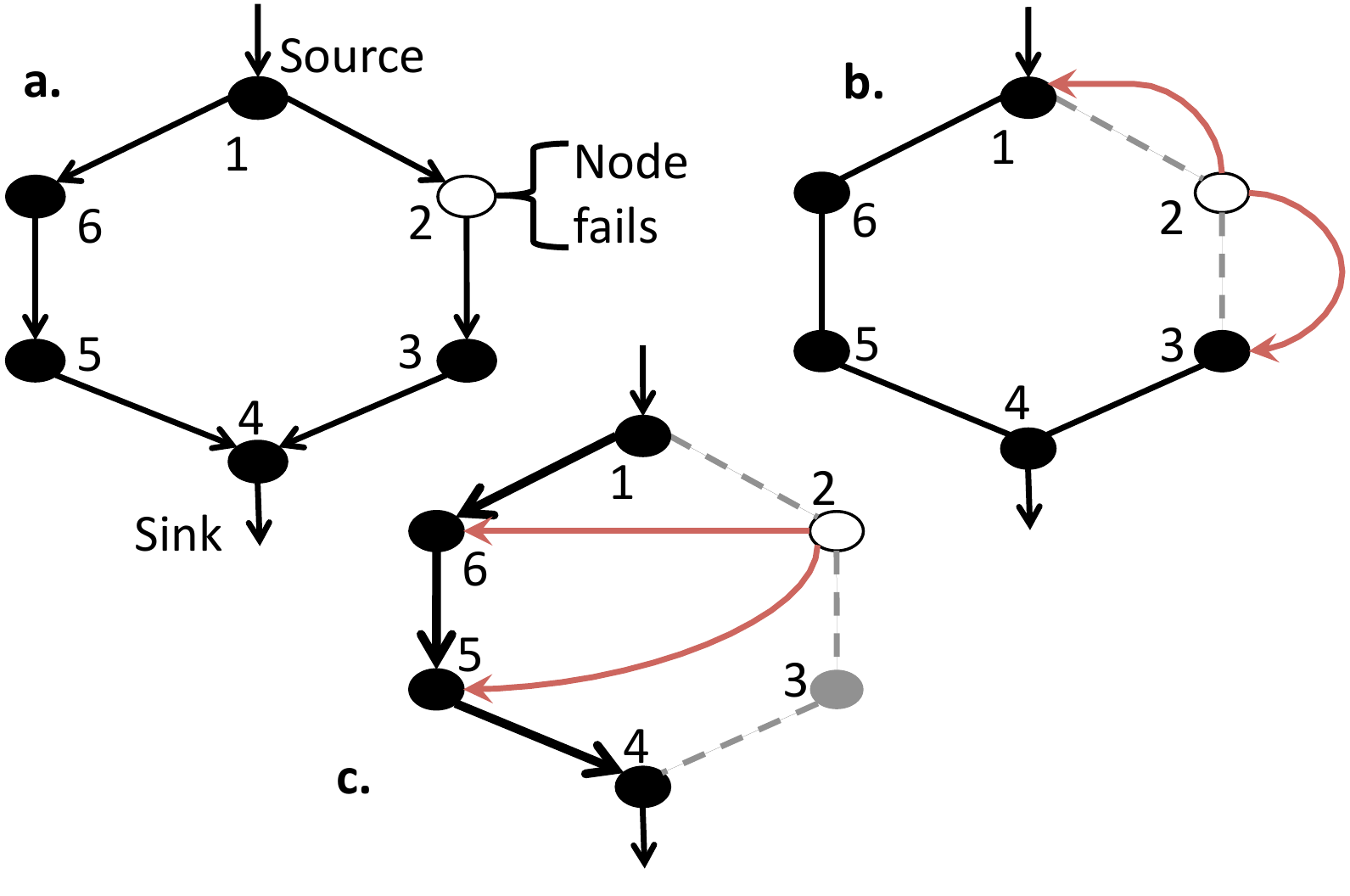}

\caption{\label{fig:cascade}An illustration of the difference between a topological
(nearest-neighbor) model of cascading failure and one based on Kirchhoff's
laws. \textbf{a.} Node 2 fails, which means that its power flow (load)
must be redistributed to functioning nodes. \textbf{b.} In many topological
models of cascading failure (e.g., \cite{Wang:2009}), load from failed
components is redistributed to nearest neighbors (Nodes 1 and 3).
\textbf{c.} In an electrical network current re-routes by Kirchhoff's
laws, which in this case means that the power that previously traveled
through Node 2 is re-routed through Nodes 5 and 6. In addition, by
Kirchhoff's laws, Node 3 ends up with no power-flow. }

\end{figure}

The value of purely topological models of power grid failure in assessing
actual failure modes in the electricity infrastructure is not well-established.
Commodity (electric energy) flows in electricity networks are governed
by Ohm's law and Kirchhoff's laws, which are not captured particularly
well in simple topological models (see Fig.\ref{fig:cascade}). Some
have identified relationships between the physical properties of power
grids and topological metrics \cite{Sole:2008,Arianos:2009,Hines:2010a},
and find that some metrics do correlate to measures of power system
performace. However, to our knowledge, no existing research has systematically
compared the results from a power-flow based cascading failure vulnerability
model with those from graph theoretic models of vulnerability. Because
cascading failures (and hurricanes) cause the largest blackouts \cite{Hines:2009}
and contribute disproportionately to overall reliability risk \cite{Dobson:2007},
models that incorporate the possibility of cascading failure are necessary
to provide a sufficiently broad view of power network vulnerability.
While there is extensive literature on cascading failure and contagion
in abstract networks (see, e.g., Sec. 4 of \cite{Boccaletti:2006}),
and some application of these methods to power networks \cite{Kinney:2005},
direct comparisons are needed to draw firm conclusions about the utility
of topological methods.

Our primary goal, therefore, is is to compare the vulnerability conclusions
that result from topological measures of network vulnerability with
those that result from a more realistic model of power network failure.
We draw on existing literature (particularly \cite{Albert:2004,Albert:2000})
to choose the topological vulnerability measures used in this paper.

\section{Vulnerability measures}

Our first vulnerability measure is \emph{characteristic path length}
($0<L<\infty$), which is the average distance among node pairs in
a graph. In \cite{Albert:2000}, path length (network diameter) was
suggested as a measure of network vulnerability because as more components
fail nodes become more distant, which may indicate that flows within
the network are inhibited.

The second measure is \emph{connectivity loss} ($0<C<1$), which was
proposed in \cite{Albert:2004} as a way to incorporate the locations
of sources (generators) and sinks (loads) into a measure of network
vulnerability. Connectivity loss is defined: $C=1-\left\langle n_{g}^{i}/n_{g}\right\rangle _{i}$,
where $n_{g}$ is the number of generators in the network and $n_{g}^{i}$
is the number of generators that can be reached by traveling from
node $i$ across non-failed links.

The third measure, which does not appear in the existing network science
literature, is \emph{blackout sizes} as calculated from a model of\emph{
}cascading failure in a power system. While a perfect model of cascading
failure would accurately represent the continuous dynamics of rotating
machines, the discrete dynamics associated with relays that disconnect
stressed components from the network, the non-linear algebraic equations
that govern flows in the network, and the social dynamics of operators
working to mitigate the effects of system stress, all power system
models simplify these dynamics to some extent. Unlike simple topological
metrics, our model does capture the effects of Ohm\textquoteright{}s
and Kirchhoff\textquoteright{}s laws, by using linear approximations
of the non-linear power flow equations \cite{Bergen:1986}. Similar
models have been used to study cascading failure in a number of recent
papers \cite{Carreras:2004,Dobson:2007,Mei:2009}. 

In our model, when a component fails, the {}``DC power-flow'' equations
are used to calculate changes in network flow patterns. In the DC
approximation the net power injected into a node (generation minus
load: $P_{i}=P_{g,i}-P_{d,i}$) is equal to the total amount of power
flowing to neighboring nodes through links (transmission lines or
transformers): $P_{i}=\sum_{j}(\theta_{i}-\theta_{j})/X_{ij}$, where
$\theta_{i}$ is the voltage phase angle at node $i$, and $X_{ij}$
is the series reactance of the link(s) between nodes $i$ and $j$.
Each link has a relay that removes it from service if its current
exceeds 50\% of its rated limit for 5 seconds or more. The trip-time
calculations are weighted such that the relays will trip faster given
greater overloads. While it is true that over-current relays are not
universally deployed in high-voltage power systems, they provide a
good approximation of other failure mechanisms that are common, such
as lines sagging into underlying vegetation (an important contributor
to the August 14, 2003 North American blackout \cite{USCA:2004}).
After a component fails the model recalculates the power flow and
advances to the time at which the next component will fail, or quits
if no further components are overloaded. If a component failure separates
the grid into unconnected sub-grids, the following process is used
to re-balance supply and demand. If the imbalance is small, such that
generators can adjust their output by not more than 10\% and arrive
at a new supply/demand balance, this balance is achieved through generator
set-point adjustments. If this adjustment is insufficient, the smallest
generator in the sub-grid is shut down until there is an excess of
load. If there is excess load after these generator adjustments, the
simulator curtails enough load to balance supply and demand. This
balancing process approximates the process that automatic controls
and operators follow to balance supply and demand during extreme events.
The size of the blackout ($S$) is reported at the end of the simulation
as the total amount of load curtailed.

\section{Attack vectors}

In order to measure power network vulnerability, we test the response
of 41 electricity networks to a variety of exogenous disturbance vectors
(attacks or random failures). In each case we measure the relationship
between disturbance (attack or random failure) size and disturbance
cost using the three vulnerability measures described above. To compare
our results with prior research five disturbance vectors are simulated.
These are described as follows.

The first vector is \emph{random failure}, in which nodes (buses)
are selected for removal by random selection, with an equal failure
probability for each node. This approach simulates failure resulting
from natural causes (e.g., storms) or an unintelligent attack. For
each network, we test its response to 20 unique sets of random failures,
with 10 nodes in each set. These sets are initially selected from
a uniform distribution, and then applied incrementally (one node,
then two nodes, etc.).

The second vector is \emph{degree attack}, in which nodes are removed
incrementally, starting with the highest degree (connectivity) nodes.
This strategy represents an intelligent attack, in which the attacker
chooses to disable nodes with a large number of neighboring nodes.

The third vector is a \emph{maximum-traffic attack}, in which nodes
are removed incrementally starting with those that transport the highest
amounts of power. We use the term {}``traffic'' to differentiate
this measure from {}``load,'' which frequently describes the quantity
of power being consumed at a node. Thus traffic ($T$) is similar
to the measures described as load in \cite{Albert:2004,Wang:2009}.
The following measure of node-loading is used to select maximum-traffic
nodes: $T_{i}=|P_{i}|+\sum_{j}|(\theta_{i}-\theta_{j})/X_{ij}|$. 

The fourth vector is \emph{minimum-traffic attack}, which is the inverse
of the max-traffic attack. This vector is used for comparison with
the conclusions in \cite{Wang:2009}, which argues that failures at
low-traffic (load) nodes lead to larger blackouts than failures at
high-traffic nodes.

The fifth vector is \emph{betweenness attack}, in which nodes are
removed incrementally, starting with those that have the highest betweenness
centrality (the number of shortest paths that pass through a node
\cite{Boccaletti:2006}). This vector was used in \cite{Albert:2004}
to approximate an attack on high traffic (load) nodes, and was reported
to result in disproportionately large failures.

\section{Results}

\begin{figure}[tbh]
\includegraphics[width=1\columnwidth,height=4.3in]{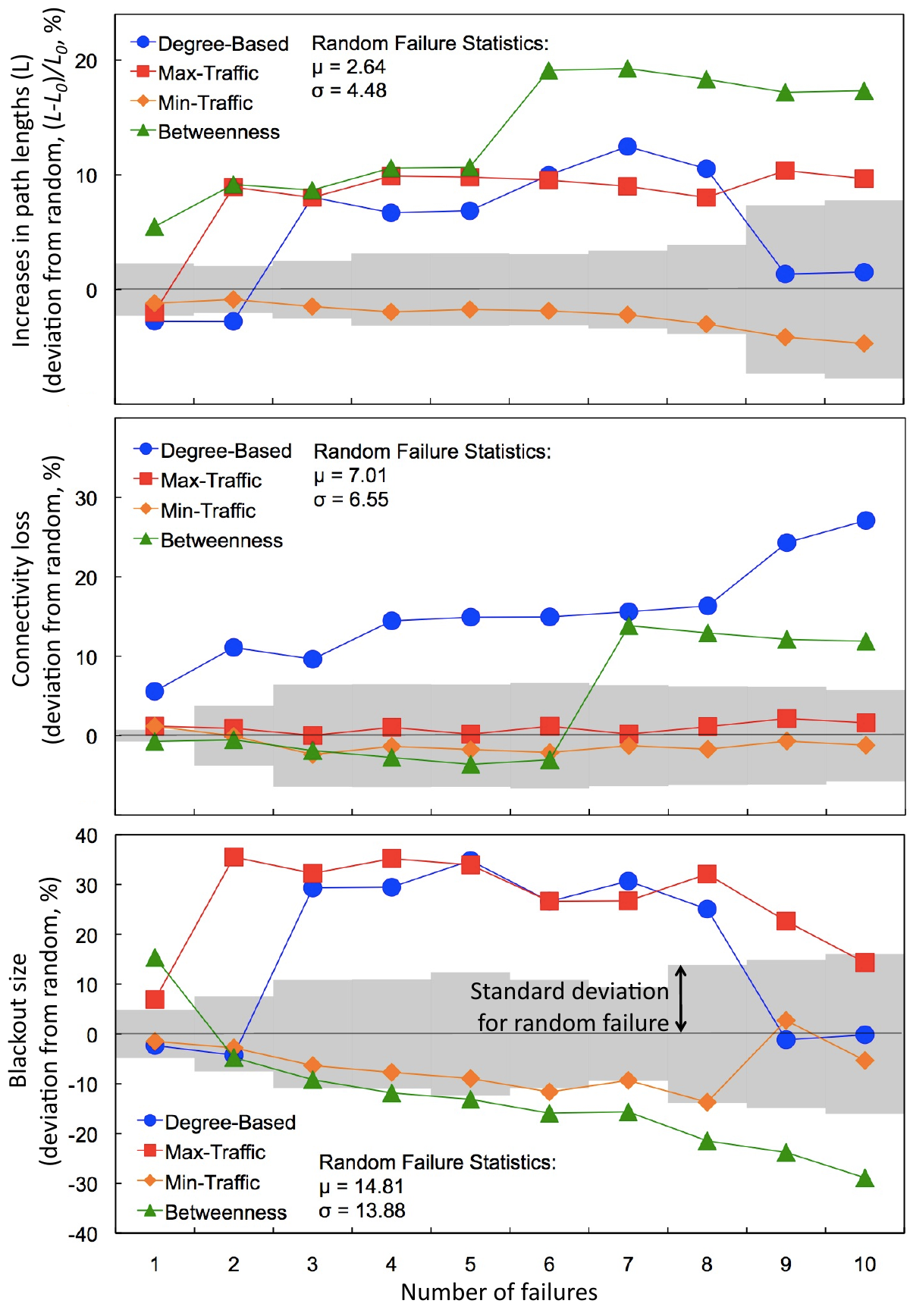}

\caption{\label{fig:IEEE300-results}(color online). Simulated response of
the IEEE 300 bus network to directed attacks. The top panel shows
the change in characteristic path lengths ($L$) as the number of
failures increases. The middle panel shows connectivity loss ($C$)
and the bottom panel shows the size of the resulting blackout both
as a function of the number of components failed. The results for
random failures are averages over 20 trials. The trajectories shown
are differences between the attack-vector results and the random failure
averages. Shading indicates $\pm1\sigma$ for the random failures. }

\end{figure}

\begin{figure}[tbh]
\includegraphics[width=1\columnwidth,height=4.3in]{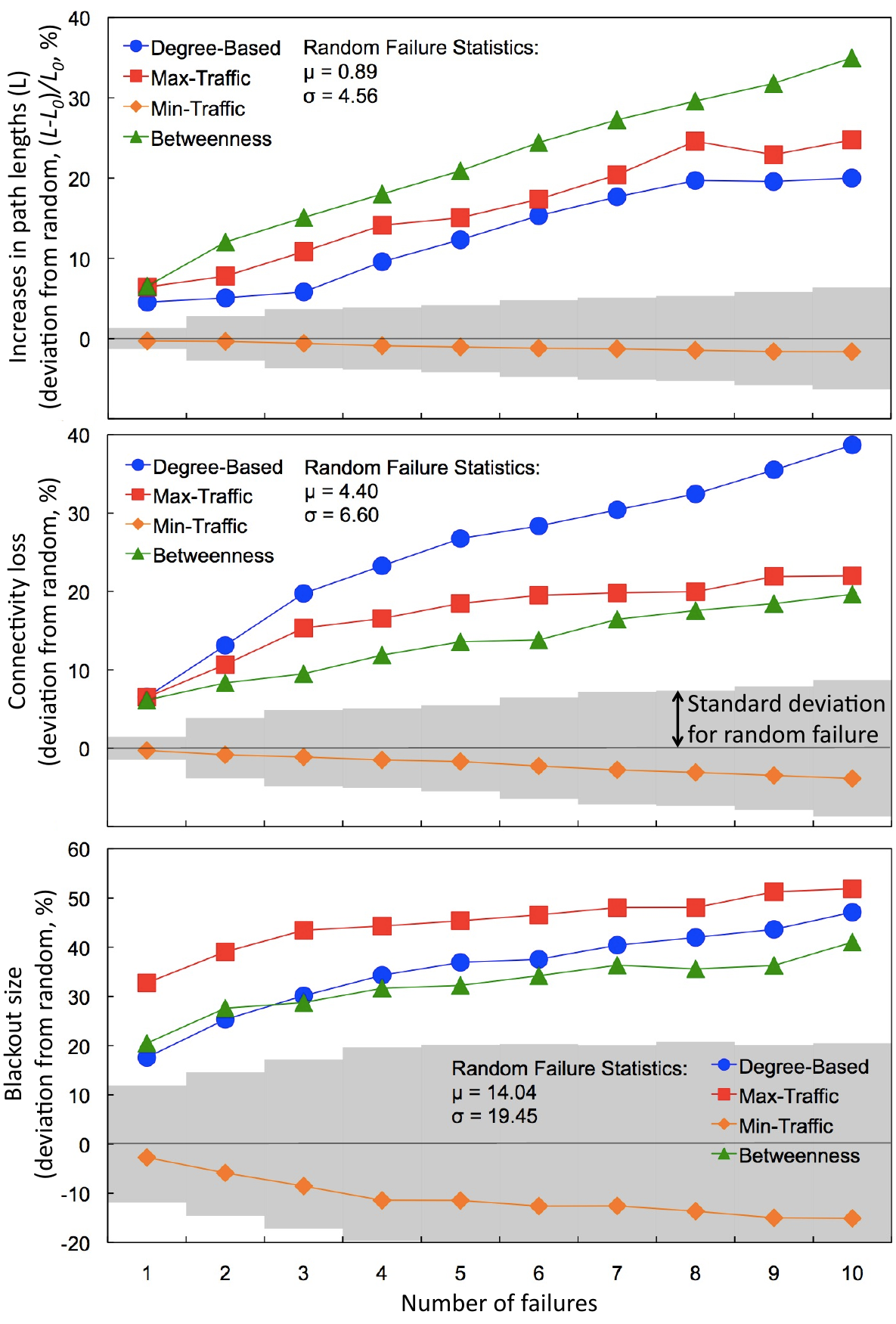}

\caption{\label{fig:EI-results}(color online). Simulated response of 40 control
areas in the Eastern Interconnect network to directed attacks. The
top panel shows the average characteristic path lengths ($L$) as
the number of failures increases. The middle panel shows connectivity
loss ($C$) and the bottom panel shows the size of the resulting blackout
both as a function of the number of components failed. The results
for random failures are averages over 20 trials in each of the 40
areas. The trajectories shown are differences between the attack-vector
results (averaged over the 40 areas) and the random failure averages.
Shading indicates $\pm1\sigma$ for the random failures. }

\end{figure}

To compare the vulnerability measures we report results from the simulation
of random failures and directed attacks for a common test system (IEEE
300 bus test case) and 40 of 136 control areas from within the North
American Eastern Interconnect (EI). The EI data come from a North
American Electric Reliability Corporation power-flow planning case,
to which the authors have been granted access for the purpose of this
research.%
\footnote{The North American power grid data used in this paper are available
from the US Federal Energy Regulatory Commission, through the Critical
Energy Infrastructure Information request process (http://www.ferc.gov/legal/ceii-foia/ceii.asp).%
} The 40 control areas analyzed were selected because of their proximate
sizes (336-1473 nodes). Together they represent 29,261 of 49,907 nodes
(buses) in the Eastern Interconnect data. We initialized the simulations
to provide an initial balance between supply and demand by decreasing
either load or generation, whichever was initially greater. In a few
areas the base-case power flows exceeded the rated flow limits. In
these cases we increased the line limits until all power flows were
10\% below the flow limits. Actual locations have been deleted from
our data set, such that these results are not linked to physical locations
in the US electricity infrastructure. 

\begin{figure}[tbh]
\includegraphics[width=1\columnwidth]{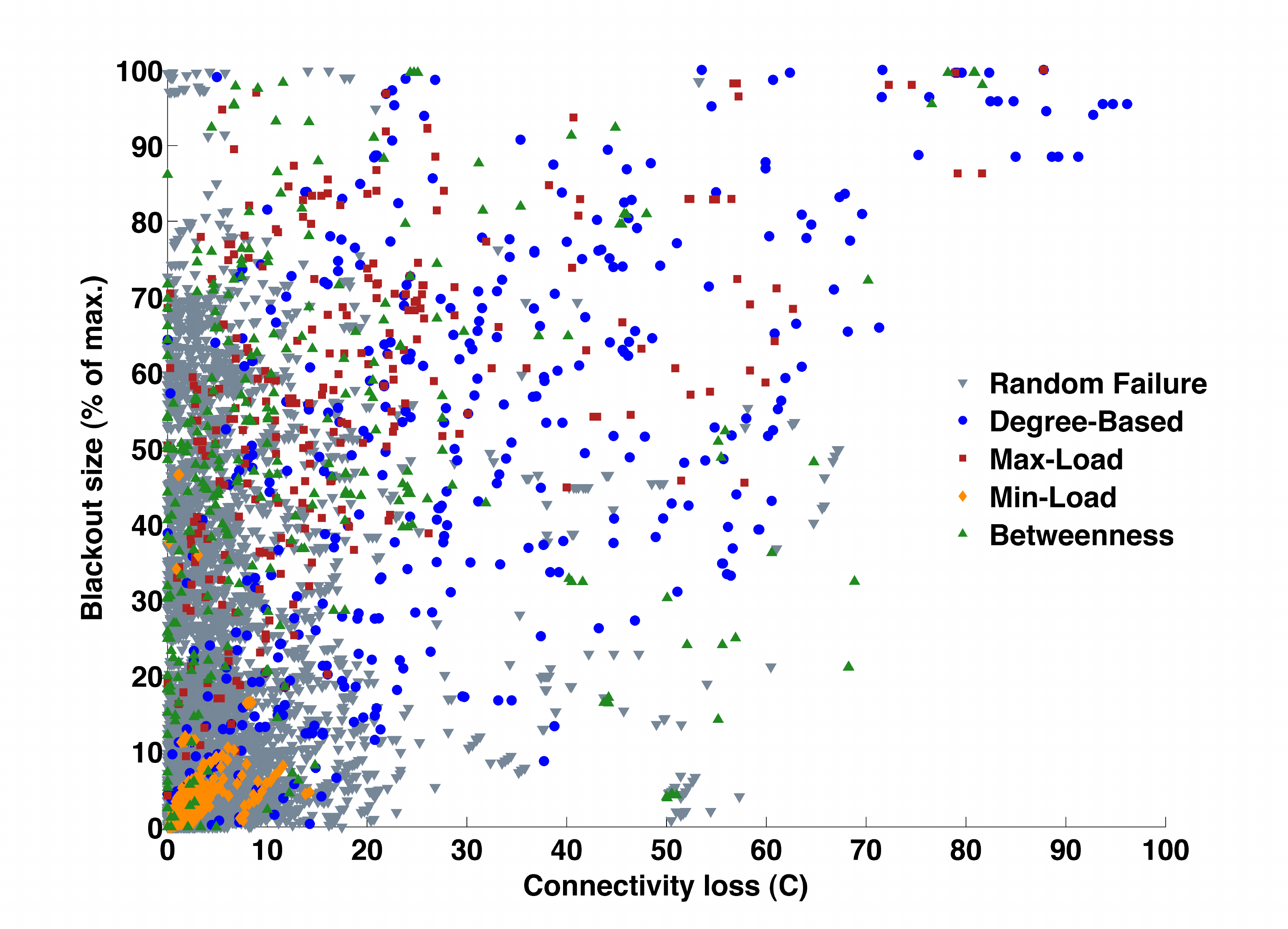}

\caption{\label{fig:CvsB}(color online). The correlation between blackout
sizes and connectivity loss ($C$) for 40 areas within the EI network.
The correlation coefficients corresponding to each attack vector are
as follows: $\rho=0.210$ (random failure), $\rho=0.621$ (degree
attack), $\rho=0.551$ (max-traffic attack), $\rho=0.288$ (min-traffic
attack), $\rho=0.138$ (betweenness attack), and $\rho=0.477$ (all
simulations).}

\end{figure}

The upper panels of Figs.\ \ref{fig:IEEE300-results} and \ref{fig:EI-results}
show how path lengths ($L$) change as nodes are removed from the
test networks. In both the IEEE 300 bus network and the EI areas path
lengths resulting from degree-based, max-traffic, and betweenness
attacks is greater than the average $L$ from random failures. Min-traffic
attacks do not substantially differ from random failures in this measure.

The middle panels of Figs. \ref{fig:IEEE300-results} and \ref{fig:EI-results}
illustrate the difference between the connectivity losses ($C$) from
directed attacks and $C$ from random failure. From this semi-topological
perspective, power grids are notably more vulnerable to directed attacks
than to random failure, and are thus similar to scale-free networks
(see \cite{Albert:2004} for a similar result).

The blackout size results (lower panels of Figs. \ref{fig:IEEE300-results}
and \ref{fig:EI-results}) also indicate that power networks are notably
more vulnerable to directed (degree-based, max-traffic- and betweenness-based)
attack than they are to random failure. Max-traffic attacks on 10
nodes produce blackouts with an average size of 72\%. Random failure
of 10 nodes results in an average blackout size of 20\%, and min-traffic
attacks produced much smaller blackouts (5\% average). From these
results it appears that the prediction in \cite{Wang:2009} that attacks
on low-traffic nodes lead to large failures is not accurate. Note
that the measure of traffic (load) used \cite{Wang:2009} is different
than ours, but it would be incorrect to conclude that failures at
low power-flow nodes contribute substantially to system vulnerability.

While trends in the path length, connectivity loss and blackout size
measures are similar after averaging over many simulations, the correlation
between measures for individual simulations is poor. Because connectivity
loss does not directly account for cascading failure, it roughly predicts
only the minimum size of the resulting blackout (see Fig. \ref{fig:CvsB}).
Once triggered, the complex interactions among network components
during a cascading event can result in a blackout of almost any size.
Many disturbances with small connectivity loss (<10\%) produced very
large blackouts.

Another notable difference among the model results is that one would
draw different conclusions about the most dangerous attack vectors,
depending on the vulnerability measure used. From path lengths, betweenness
attacks have the greatest impact. From connectivity loss, degree-based
attacks look most dangerous. From the blackout model, max-traffic
attacks appear to contribute most to vulnerability.

\section{Conclusions}

Together these results indicate that while topological measures can
provide some indication of general vulnerability trends, they can
also be misleading when used in isolation. In some cases, overly-abstracted
topological models can result in erroneous conclusions, which could
lead to mis-allocation of risk-mitigation resources. Vulnerability
measures that properly account for network behavior as well as the
arrangement of sources and sinks produce substantially different results;
we argue that these results are more realistic and more useful for
infrastructure risk assessment. If the results described here are
similar to what one would obtain from an ideal model of cascading
failure, the implication for electricity infrastructure protection
is that the defense of high-traffic, high-degree, and high-betweenness
substations from attack is likely to be a cost-effective risk mitigation
strategy.

\section*{Acknowledgement}

This work is supported in part by the US National Science Foundation,
Award \#0848247.

\bibliographystyle{IEEEtran}
\bibliography{EnergySystemsGroup}

\end{document}